\documentclass[aps,prl,twocolumn,showpacs,floatfix]{revtex4}
\usepackage{graphicx,bm,amsmath,amssymb,natbib,url,epsfig,psfrag}
\usepackage{color}

\begin{document}

\title{Bilayer Graphene Interferometry: Phase Jump and Wave Collimation}
\author{Sunghun Park}
\author{H.-S. Sim}
\affiliation{Department of Physics, Korea Advanced Institute of
Science and Technology, Daejeon 305-701, Korea}

\date{\today}

\begin{abstract}
We theoretically study the phase of the reflection amplitude of an electron (massive Dirac fermion)
at a lateral potential step
in Bernal-stacked bilayer graphene.
The phase shows anomalous jump of $\pi$, as the electron incidence angle (relative
to the normal direction to the step) varies to pass
$\pm \pi/4$.
The jump is attributed to the Berry phase associated with the pseudospin-$1/2$ of the electron.
This Berry-phase effect is robust against the band gap opening due to the external
electric gates generating the step.
We propose an interferometry setup in which collimated waves
can be generated and tuned.
By using the setup, one can identify both the $\pi$ jump and the collimation angle.
\end{abstract}



\maketitle


{\it Introduction.---} Monolayer graphene has attracted much attention due to
potential application of its unusual properties~\cite{NetoRMP}.
Its low-energy quasiparticles
are massless Dirac fermions. They show Klein tunneling~\cite{Katsnelson},
anomalous quantum Hall effects~\cite{Novoselov1, Zhang},
electron optics behavior such as focusing~\cite{CheianovLens}
and collimation~\cite{Cheianov, GarciaPomar, CHPark, Pereira}, etc.
The Klein tunneling
is understood by the chirality of the quasiparticles~\cite{Katsnelson},
or equivalently by Berry phase~\cite{Ando, Shytov}.
There have been experimental efforts to observe it
in a bipolar junction~\cite{Gordon, Gorbachev} and
in an interferometry~\cite{Young}.

Bilayer graphene has properties quite different from the monolayer.
For example, in Bernal-stacked bilayer, low-energy
quasiparticles are massive Dirac fermions~\cite{McCann}.
There have been studies on bilayer-graphene Klein effects~\cite{Katsnelson},
quantum Hall effects~\cite{Novoselov2},
band-gap engineering~\cite{McCann, Ohta, Oostinga, Castro, SanJose},
etc.
However,
more studies on the bilayer may be necessary for the understanding
of the features of the massive Dirac fermions such as their transport related
with Berry phase~\cite{Novoselov2, Mikitik} and collimation~\cite{Katsnelson}.

\begin{figure}
\includegraphics[width=0.42\textwidth]{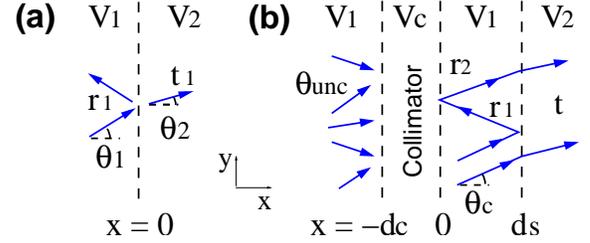}
\caption{(Color online)
(a) Lateral monopolar potential step in Bernal-stacked bilayer graphene ($xy$ plane).
The arrows represent the propagation of electron waves, the dashed lines
show the boundaries between the regions with different potential strengths $V_i$'s,
$\theta_{1(2)}$ is the propagation angle of the incident (transmitted) wave,
and $r_i$'s ($t_i$'s) are the reflection (transmission) amplitudes.
As $\theta_1$ varies, the phase of $r_1$ can exhibit an abrupt jump of $\pi$ at $\theta_1 = \pm \pi /4$,
a Berry-phase effect.
(b) Interferometry setup for the detection of the $\pi$ phase jump of $r_1$,
based on the interference between the two paths drawn in $x \in [0, d_s]$.
To see the incidence-angle (here, $\theta_c$) dependence of $r_1$,
collimated waves are generated from a collection of waves with different
incidence angle $\theta_\textrm{unc}$, by using the
resonant filtering due to the barrier $V_c$.
The collimation angle $\theta_c$ can be also identified, using the interference.
}
\label{fig1:Setup}
\end{figure}

In this Letter, we theoretically study the reflection of a low-energy electron
at a lateral potential step of monopolar ($p$-$p$ or $n$-$n$) type
in Bernal-stacked bilayer graphene [Fig.~\ref{fig1:Setup}(a)].
The phase of the reflection amplitude is found to show anomalous behavior, as the
electron incidence angle $\theta_1$ (relative to the normal direction to the step) varies.
It shows an abrupt jump of $\pi$ at $\theta_1 = \pm \pi / 4$ when the step height is
much smaller than the kinetic energy of the electron.
Based on a reversal symmetry [Eq.~\eqref{Toperator}],
we attribute the $\pi$ jump to the Berry phase associated with
the pseudospin-1/2 of the electron.
The jump becomes gradual, as the step height increases,
due to the evanescent waves existing at the step boundary
and breaking the reversal symmetry.
We show that the phase jump can be detected in an interferometry setup
[Fig.~\ref{fig1:Setup}(b)].
We remark that the phase jump is robust against the band gap opening due to
the electric gates generating the step,
and that
the setup does not require an external magnetic field,
contrary to previous works~\cite{Novoselov1, Shytov, Zhang, Young, Novoselov2}
for the detection of Berry phase effects in graphene.

{\it Potential step.---}
In Bernal-stacked bilayer graphene, a low-energy electron states $\Psi_\textrm{K} (x,y)$
with energy $E$ in the K valley is governed by Hamiltonian~\cite{McCann}
$H_\textrm{K} \Psi_\textrm{K} = E \Psi_\textrm{K}$,
\begin{eqnarray}
H_\textrm{K} & = &  \frac{v^2}{\gamma} \vec{\sigma}_\textrm{K} \cdot \vec{q} + V (x), \,
\,\,
\vec{q} \equiv ( -p^{2}_{x} + p^{2}_{y}, - 2 p_{x} p_{y}). \label{Qvector}
\end{eqnarray}
The two pseudospin components of $\Psi_\textrm{K}$
describe
the lattice sites A$_1$ and B$_2$ of the bilayer,
where A$_l$ and B$_l$ denote the two basis sites of layer $l=1,2$.
$\vec{\sigma}_K = (\sigma_x, \sigma_y)$ is the pseudospin operator
for the K valley,
$\vec{p} = (p_x, p_y)$ is the momentum
relative to the valley center,
$\gamma \approx 0.39 \, \, e\textrm{V}$
is the interlayer coupling,
$v \approx 10^6$~m/s,
and
$\sigma_{x, y, z}$ are the Pauli matrices.
%
%
$H_\textrm{K}$
is valid for
$0.002 \gamma < |E-V| \ll \gamma$,
where the trigonal warping~\cite{McCann} is ignored.
We study the K valley only, as the K$'$ valley shows the same result,
and ignore the intervalley mixing as $V(x)$ slowly varies on
the scales of the lattice constant and
$\hbar v / \gamma$.
Band gap~\cite{McCann} due to the
gates generating $V(x)$ 
will be considered later. 

It is worthwhile to see a symmetry
of $H_\textrm{K}$, which has not been 
discussed in literatures.
$H_\textrm{K}$ is invariant under the antiunitary
operator $\Theta$ (for $0.002 \gamma < |E-V| \ll \gamma$),
\begin{eqnarray}
\Theta & = & i \sigma_y \mathcal{R}_{\vec{p},\pm \pi / 2} \mathcal{C},
\label{Toperator}
\end{eqnarray}
when $\vec{p}$ is real (which is achieved when
$V_1 - V_2 \to 0$ in Fig.~\ref{fig1:Setup}).
Here, $\mathcal{R}_{\vec{p}, \pm \pi/2}$ is the operator rotating
$\vec{p}$ by angle $\pm \pi / 2$ [i.e., $(p_x, p_y) \to
( \pm p_y, \mp p_x )$ and thus $\vec{q} \to - \vec{q}$], $i \sigma_y \mathcal{C}$
reverses pseudospin, and $\mathcal{C}$ is the complex conjugation operator.
We call this invariance as reversal symmetry hereafter, in the sense that
$\Theta$ is exactly the same as the time reversal operator defined for a single valley~\cite{Beenakker}
if $\vec{q}$ is replaced by $\vec{p}$ in Eq.~\eqref{Qvector}.
The symmetry $\Theta$ and the Berry phase $\pi$~\cite{NOTE2}, associated with
a loop encircling once the origin in the $\vec{q}$ space, can give rise to
an interesting effect in a monopolar potential step
(see below).
This effect corresponds to the Klein tunneling in bipolar monolayer graphene, which
results from the time reversal symmetry and the Berry phase $\pi$ due to a $\vec{p}$-space
loop~\cite{Shytov, Ando}.

%

\begin{figure}
\includegraphics[width=0.45\textwidth]{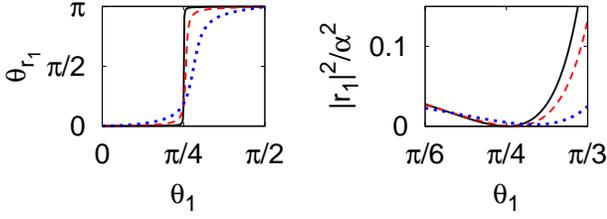}
\caption{(Color online)
Dependence of the reflection phase $\theta_{r_1}$ and
probability $|r_1|^2$ (scaled by $\alpha^2$) on $\theta_1$
for $\alpha =$ $- 0.01$ (solid), $- 0.1$ (dashed), and $- 0.5$ (dotted).
Around $\theta_1 = \pi/4$,
$\theta_{r_1}$ shows a jump of $\pi$, accompanied by $|r_1|^2 = 0$.
}
\label{fig2:Rephase}
\end{figure}

Now, we consider a potential step $V(x)$ of monopolar type,
where $V(x) = V_1$ for $x< 0$ and $V_2$ for $x > 0$,
and study the reflection amplitude $r_1$ of a plane wave
incoming to the step
with energy $E$ and incidence angle $\theta_1$ [Fig.~\ref{fig1:Setup}(a)],
where $E > V_1, V_2$ ($E < V_1, V_2$) for a $n$-$n$ ($p$-$p$) step.
Due to the translational invariance along $\hat{y}$,
the wave is described by
$\Psi_{\textrm{K}, k_y} = e^{i k_y y} \psi_{\textrm{K}, k_y} (x)$.
$\psi_{\textrm{K}, k_y} (x)$ is a superposition of
propagating and evanescent waves~\cite{Katsnelson},
\begin{eqnarray} 
& & e^{i k_{1_x} x}
\left(
    \begin{array}{c}
      e^{- i \theta_1} \\
      - s_1 e^{i \theta_1}
    \end{array}
\right)
    + r_1 e^{- i k_{1_x} x}
\left(
    \begin{array}{c}
      - e^{i \theta_1} \\
      s_1 e^{- i \theta_1}
    \end{array}
\right) \nonumber\\
& + & a e^{\kappa_1 x}
\left(
    \begin{array}{c}
      \sqrt{1 + \textrm{sin}^2 \theta_1} + s_1 \textrm{sin} \theta_1 \\
      s_1 \sqrt{1 + \textrm{sin}^2 \theta_1} - \textrm{sin} \theta_1
    \end{array}
\right) \,\,\,\, \textrm{for $x < 0$, } \,\,\, \nonumber
\end{eqnarray}
\begin{eqnarray} 
& & t_1 e^{i k_{2_x} x}
\left(
    \begin{array}{c}
      e^{- i \theta_2} \\
      - s_2 e^{i \theta_2}
    \end{array}
\right) \nonumber \\
& + & b e^{- \kappa_2 x}
\left(
    \begin{array}{c}
       - \sqrt{1 + \textrm{sin}^2 \theta_2} + s_2 \textrm{sin} \theta_2 \\
       - s_2 \sqrt{1 + \textrm{sin}^2 \theta_2} - \textrm{sin} \theta_2
    \end{array}
\right) \,\,\,\, \textrm{for $x > 0$. } \,\,\, \nonumber
\end{eqnarray}
Here, $s_j = \textrm{sgn} (E - V_j)$,
$\hbar k_y = s_1 \sqrt{\vert E - V_1 \vert \gamma / v^2}\ \textrm{sin} \theta_1 $,
$\hbar k_{j_x} = s_j \sqrt{\vert E - V_j \vert \gamma / v^2}\ \textrm{cos} \theta_j$,
$\hbar \kappa_j = \sqrt{\vert E - V_j \vert \gamma / v^2 } \sqrt{1 + \textrm{sin}^2\theta_j }$,
$j \in \{ 1, 2 \}$
refers to the region with $V_j$,
and the propagation angle $\theta_2$ of the transmitted wave is governed by the conservation
of $p_y$, $s_1 \sqrt{|E - V_1|}\ \textrm{sin} \theta_1 = s_2 \sqrt{|E - V_2|}\ \textrm{sin} \theta_2$.
The coefficients $r_1$, $t_1$, $a$, and $b$ are determined by the continuity of
$\psi_{\textrm{K},k_y}(x)$ and $d \psi_{\textrm{K},k_y}(x)/dx$ at $x=0$.
We introduce a parameter $\alpha \equiv (V_2 - V_1) / (E - V_1)$,
the ratio of step height and the kinetic energy of the incident wave.
The following effects depend only on $\alpha$ and $\theta_1$,
regardless of other details such as the type ($n$-$n$ or $p$-$p$) of the step.

%

\begin{figure}
\includegraphics[width=0.45\textwidth]{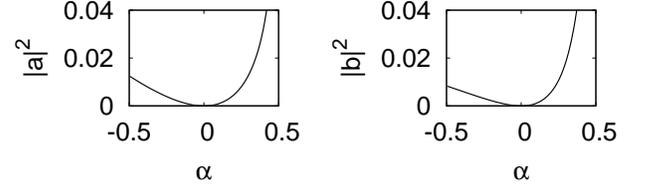}
\caption{
Dependence of the probabilities $|a|^2$ and $|b|^2$ of
the evanescent waves on $\alpha$ at $\theta_1 = \pi / 4$.}
\label{fig3:Locwave}
\end{figure}

{\it Phase jump and Berry phase.---}
We examine the dependence of the reflection amplitude $r_1 = |r_1| e^{i \theta_{r_1}}$
on $\alpha$ and $\theta_1 \in [0, \pi/2]$; $r_1$ is an even function of $\theta_1$.
We first discuss the case of
$|\alpha| \ll 1$, and then that of finite $\alpha$.

For $\vert \alpha \vert \ll 1$, the reflection phase $\theta_{r_1}$ shows
an abrupt jump of $\pi$ at $\theta_1 = \pi/4$; see Fig.~\ref{fig2:Rephase}.
The phase jump is accompanied by a reflection zero, $|r_1|^2 = 0$.
To see the behavior, we derive
the expression of $r_1$ near $\theta_1 = \pi/4$,
\begin{eqnarray}
r_1 \simeq - \frac{\textrm{cos} 2 \theta_1}{4 \textrm{cos}^{2} \theta_1}\ \alpha
\,\,\,\,\,\,\,\,\, \textrm{for } |\alpha| \ll 1 \textrm{ and } \theta_1 \simeq \pi/4.
\label{Resoln}
\end{eqnarray}
The $\pi$ jump occurs,
irrespective of the sign of $\alpha$.
And, for $\vert \alpha \vert \ll 1$, the evanescent waves
are ignorable ($|a|^2$, $|b|^2 \simeq 0$),
as shown for $\theta_1 = \pi / 4$ in Fig.~\ref{fig3:Locwave}; the same
occurs for other values of $\theta_1$.
Below, we attribute the $\pi$ jump to the Berry phase associated with the pseudospin.

\begin{figure}
\includegraphics[width=0.3\textwidth]{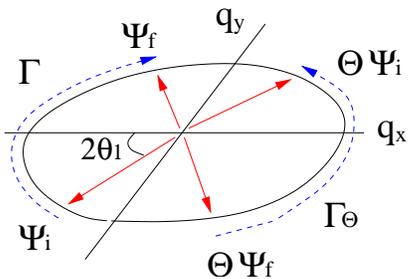}
\caption{(Color online)
Schematic view of the change of $\vec{q}$ in the reflection process for
$|\alpha| \ll 1$. This view also shows the change of
pseudospin, since pseudospin is parallel (antiparallel) to $\vec{q}$ in
the $n$-$n$ ($p$-$p$) step.
The change from an incident wave $\Psi_i$ (with $\theta_1 < \pi / 4$)
to its reflected wave $\Psi_f$ is represented by $\Gamma$. The change of
the reversal-symmetric process from $\Theta \Psi_f$ (which incidence angle $> \pi/4$)
to $\Theta \Psi_i$ is denoted by $\Gamma_\Theta$.
At $\theta_1 = \pi / 4$, $\Psi_i$ ($\Psi_f$) has the same $\vec{q}$ as
$\Theta \Psi_f$ ($\Theta \Psi_i$), but
the change of the pseudospin in the process $\Gamma$ differs
from that in $\Gamma_\Theta$ by $2 \pi $ rotation,
resulting in the abrupt $\pi$ jump
(Berry phase $\pi$ in the $\vec{q}$ space) of the reflection phase $\theta_{r_1}$.
}
\label{fig4:Qspace}
\end{figure}

According to
the Hamiltonian~\eqref{Qvector},
the pseudospin couples with the vector
$\vec{q}$ such that
it is parallel (antiparallel) to $\vec{q}$ in the $n$-$n$ ($p$-$p$) step.
We follow the change of $\vec{q}$,
to see the change of the pseudospin in the reflection.
The vectors $\vec{q}$ of
an incident wave $\Psi_i$ with $\theta_1 = \pi / 4 - \delta$
(where $\delta$ is a small positive angle)
and the wave $\Psi_f$ formed by the reflection of $\Psi_i$ at the step
are drawn in Fig.~\ref{fig4:Qspace}.
One can assign a clockwise path $\Gamma$ to the change of $\vec{q}$;
the choice between clockwise and counterclockwise does not matter.
We also follow the change in the reversal-symmetric process $\Gamma_\Theta$
from $\Theta \Psi_f$ to $\Theta \Psi_i$.
This process is also a solution of Eq.~\eqref{Qvector} with the same energy,
since the evanescent waves can be ignored and thus $[H_\textrm{K}, \Theta] = 0$.
Notice that the incidence angle of $\Theta \Psi_f$ is $\pi/4 + \delta$,
and that
$\Gamma$ and $\Gamma_\Theta$ are reflection-symmetric about $q_y$-axis.
As $\theta_1 \to \pi / 4$ ($\delta \to 0$), the spatial propagation in the
process $\Gamma$ becomes identical to that in $\Gamma_\Theta$.
However, at $\theta_1 = \pi / 4$,
the change of the pseudospin in the $\Gamma$ process differs from that in $\Gamma_\Theta$ by
$2 \pi$ rotation, i.e., the difference $\Gamma - \Gamma_\Theta$ forms
a loop encircling once the origin of the $\vec{q}$ space.
The resulting Berry phase $\pi$ gives rise to the abrupt jump in $\theta_{r_1}$.

We next discuss the case of finite $\alpha$.
As $|\alpha|$ increases, the abrupt jump of $\theta_{r_1}$ at $\theta_1 = \pi/4$
becomes gradual [Fig.~\ref{fig2:Rephase}].
This
is due to the evanescent waves localized at the step boundary
$x=0$, which become to affect the reflection as $|\alpha|$ increases
[Fig.~\ref{fig3:Locwave}], and
break the reversal symmetry $\Theta$; for example, a
wave $\Theta |\pm i \kappa_j, k_y \rangle \propto |k_y, \pm i \kappa_j \rangle$,
reversal-symmetric to an evanescent wave $|\pm i \kappa_j, k_y \rangle$,
is physically meaningless, as it diverges
in $\hat{y}$ direction.
For $\theta_1 \simeq \pi / 4$ and small $|\alpha|$,
we derive
\begin{eqnarray}
\frac{d\theta_{r_1}}{d\theta_1} \simeq
\frac{-(\sqrt{3} \alpha / 12)}{(\theta_1 - \pi / 4 + \alpha / 4)^2+(\sqrt{3} \alpha / 12)^2}
+ O(\alpha^3),
\label{Mrephase}
\end{eqnarray}
which shows that
the (gradual)
jump of $\theta_{r_1}$ occurs around a shifted angle
$\theta_1 = (\pi - \alpha) / 4$ within $\sqrt{3} |\alpha| / 12$. 
For $|\alpha| \gtrsim 1$, $\theta_{r_1}$ increases only gradually.
Any bipolar step does not show the phase jump, as it has $|\alpha| \ge 1$.



\begin{figure}
\includegraphics[width=0.42\textwidth]{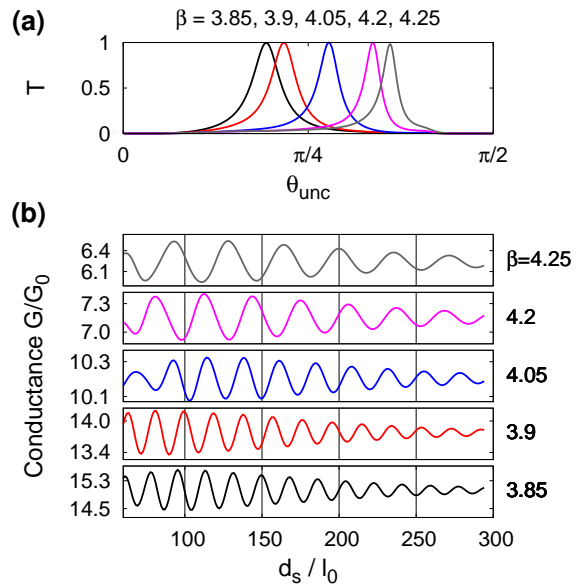}
\caption{(Color online)
(a) Transmission $T = |t|^2$ of a plane wave with incidence angle $\theta_\textrm{unc}$
through the setup in Fig.~\protect\ref{fig1:Setup}(b), for different $\beta$'s. $T$ is an even
function of $\theta_\textrm{unc}$.
(b) Conductance $G$ as a function of $d_s$.
In (a) and (b),
$E - V_1= 0.04 \gamma$, $\alpha = -0.5$, and $d_c = 60 l_0$ are chosen, while
$d_s = 60 l_0$ only in (a).
Here, $G_0 \equiv e^2 W/ (2 \pi h d_c)$
and $l_0 \equiv \hbar v / \gamma \approx 1.7 \, \, \textrm{nm}$.
}
\label{fig5:TandG}
\end{figure}

{\it Interferometry.---}
We propose an interferometry setup
for the detection of the phase jump in $\theta_{r_1}$ [Fig.~\ref{fig1:Setup}(b)].
In addition to the step,
it has a potential barrier $V_c$, which
shows transmission resonance only around some incidence angle $\theta_c$ [Fig.~\ref{fig5:TandG}(a)].
This
filtering or
collimation~\cite{Katsnelson} is used in our setup, to see the dependence of $\theta_{r_1}$ on
the incidence angle (now on the collimation angle $\theta_c$).

We first discuss the collimation.
Figure~\ref{fig5:TandG}(a) shows
the dependence of the transmission probability $T = |t|^2$ through the setup
on the incidence angle $\theta_\textrm{unc}$
of a plane wave. Here, $t$ is calculated in the same way as for the step,
and we choose $\beta \equiv (V_c - V_1) / (E - V_1) \ge 2$,
for which there is no total reflection by the barrier for any $\theta_\textrm{unc}$.
The collimation angle $\theta_c$, at which $T$ shows a maximum value for a given $\beta$,
is governed by the resonance condition, $2 k_{c_x} d_c + \varphi_0 = 2 \pi n$ ($n$ is
an integer),
where
$k_{c_x} =  s_c k_\textrm{in} \sqrt{|\beta - 1| - \sin^2 \theta_c}$ is
the $\hat{x}$-axis wave vector inside the barrier,
$k_\textrm{in} = \sqrt{|E - V_1| \gamma / (\hbar v)^2}$
is the wave vector of the incident wave,
$d_c$ is the barrier width,
$s_c = \textrm{sgn} (E-V_c)$,
and $\varphi_0$
is the reflection phase at the barrier boundaries;
in our parameter range,
the dependence of $\varphi_0$ on $\beta$
and $\theta_\textrm{unc}$ is ignorable.
The collimation is mainly done in the barrier,
and negligibly affected by the step,
as $|r_1|$ is small.
One can tune $\theta_c$ by changing $\beta$.

%
%

Due to the collimation, a wave incoming into the step has incidence angle around $\theta_c$.
Its two propagation paths in $x \in [0, d_s]$,
one with direct transmission
and
the other with reflection once at $x = d_s$ [Fig.~\ref{fig1:Setup}(b)],
result in the interference pattern of $\cos ( 2 k_{1_x} d_s + \theta_{r_1} + \theta_{r_2} )$
as a function of $d_s$,
where $\hbar k_{1_x} = s_1 k_\textrm{in} \cos \theta_c$
and $\theta_{r_2}$ is
the reflection phase at $x = 0$~\cite{NOTE1};
for $\beta \ge 2$, $\theta_{r_2}$
changes only gradually.
From the period $\lambda \equiv \pi / k_{1_x}$
and the phase shift of the pattern, one can identify
$\theta_c$ and the $\pi$ jump of $\theta_{r_1}$.


%
%

\begin{figure}
\includegraphics[width=0.45\textwidth]{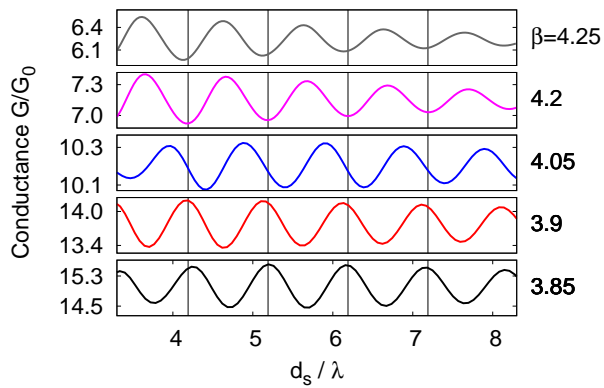}
\caption{(Color online)
The same as in Fig.~\ref{fig5:TandG}(b), but as a function of $d_s / \lambda$.}
\label{fig6:RescaledG}
\end{figure}

Based on the above idea, we analyze the conductance
$G = (4e^2/h) (W/2 \pi) \int d k_y T(k_y)$ through the setup at zero temperature and zero bias,
where the factor 4 reflects the spin and valley degeneracy,
the integral is done over all the incoming waves with the same energy $E$
but different $\theta_\textrm{unc}$,
and $W$ is the transverse width of the setup.
For different $\beta$'s and thus $\theta_c$'s,
the interference pattern in $G$ is shown in Fig.~\ref{fig5:TandG}(b), as a function of $d_s$.
The period of the pattern agrees very well with
$\lambda = \pi / k_{1_x}$, therefore,
from the period, one can identify the collimation angle $\theta_c$ in experiments, provided
that $k_\textrm{in}$ is known.
And, to see the $\pi$ jump of $\theta_{r_1}$,
we redraw $G$ in Fig.~\ref{fig6:RescaledG}, as a function of
$d_s / \lambda$; the jump is not clearly shown in Fig.~\ref{fig5:TandG}(b),
since $\lambda$ varies with $\beta$.
It shows the phase shift by $\pi$ around $\beta = 4.05$,
where $\theta_c = 1.11 \pi/4$; the value $1.11 \pi / 4$ deviates from $\pi / 4$,
and the deviation roughly agrees with the deviation $\alpha / 4$ for small $\alpha$
[see Eq.~\eqref{Mrephase}]. In this way, by tuning $\beta$ and thus $\theta_c$, one can
observe the $\pi$ jump of $\theta_{r_1}$.

%
%


{\it Discussion.---} We include the
band gap effect $\Delta(x) = \Delta_i$
due to the external gates creating $V_i$.
It is described by Hamiltonian
$H_\textrm{K}' \simeq H_\textrm{K} + \Delta (x) \sigma_z / 2$ for $\Delta \ll \gamma$,
and the argument in Fig.~\ref{fig4:Qspace} has to be modified
since $[ H_\textrm{K}', \Theta] \ne 0$ for $\Delta \ne 0$.
According to Ref.~\cite{McCann},
$\Delta_i$ can be expressed, using typical experimental parameters, as
$\Delta_i \simeq - 2 \xi_i V_i$,
where $\xi_i (\epsilon_i)
= 0.5 (1 + |\epsilon_i|) / [1 + |\epsilon_i| + \epsilon_i^2 - 0.5 \ln |\epsilon_i|]$
and $\epsilon_i \equiv V_i / \gamma$.
For small $\alpha$ [$\simeq (\epsilon_1 - \epsilon_2)/\epsilon_1$],
we approximately use constant $\xi_{i=1,2} = \xi$,
since $\xi_i (\epsilon_i)$ varies slowly enough;
we later consider the dependence of $\xi_i$ on $\epsilon_i$.
And, in the limit of zero bias and zero temperature, $E \simeq 0$.
In this regime of our interest,
one can find the unitary pseudospin-rotation $U$ satisfying
$U H_\textrm{K}' U^\dagger = (v^2/\gamma) \vec{\sigma}_\textrm{K} \cdot \vec{Q} + V(x)$,
where $\vec{Q} \equiv (\vec{q}/|\vec{q}|) \sqrt{q^2 + \gamma^2 \Delta_i^2 /(2 v^2)^2}
= \vec{q} / \sqrt{1 - \xi^2}$ for all $i$'s.
Notice that $U H_\textrm{K}' U^\dagger$ has the same form as $H_K$
and that $[U H_\textrm{K}' U^\dagger, \Theta] =0$.
Then the argument in Fig.~\ref{fig4:Qspace} is applicable
for $U \Psi_{i,f}$ and $\Theta U \Psi_{i,f}$ in the $\vec{Q}$ space,
thus, 
the abrupt $\pi$ jump in $\theta_{r_1}$ is maintained
and can be detected as in Figs.~\ref{fig5:TandG} and~\ref{fig6:RescaledG};
Eqs.~\eqref{Resoln}-\eqref{Mrephase} and Figs.~\ref{fig2:Rephase}-\ref{fig3:Locwave}
are not altered, and the dependence of $\theta_c$ on $\beta$ is modified slightly.
These features persist when we go beyond the approximation made above.
For this case of
$\xi_1 \ne \xi_2$,
Eq.~\eqref{Resoln} is shifted as $r_1 \to r_1 + \delta r_1$,
where $\delta r_1 \simeq - i \tilde{\alpha} \sin 2 \theta_1 / (4 \cos^2 \theta_1)$
and $\tilde{\alpha} = \Delta_1/[2(E-V_1)] - \Delta_2 / [2(E-V_2)]$.
From $\xi_i (\epsilon_i)$,
we find that $\delta r_1$ is ignorable
($|\tilde{\alpha}/\alpha| \lesssim 0.1$)
in reasonable ranges
of $|\epsilon_i| < 0.2$, $|\alpha| < 0.5$, and $|E/V_2| < 0.2$.
This estimation of $\delta r_1$ will be modified only slightly
when two (top and bottom) gates are used to create the step with small $V_1 - V_2$;
in this general case, the above expression of $\xi_i (\epsilon_i)$ may be altered,
but the expression of $\delta r_1$ is still valid.
Thus the $\pi$ jump in $\theta_{r_1}$ is detectable in the presence of the gap.
We emphasize the central role of the $\vec{Q}$ space as the parameter space for the Berry phase $\pi$.

We compare our result with the Klein effect in monolayer graphene.
In the monolayer, it was predicted~\cite{Shytov} and
observed~\cite{Young} in an interferometry that a sign change
($\pi$ phase jump) occurs in the back-reflection amplitude.
Contrary to our bilayer case, (i) its origin is the Berry phase $\pi$ in the $\vec{p}$ space,
(ii) it occurs at zero incidence angle, thus an external magnetic field may be required to
detect it,
and (iii) it occurs in a bipolar junction.


In summary, we find
the abrupt jump $\pi$ of the reflection phase at a monopolar
potential
step in Bernal-stacked bilayer graphene.
The jump is the manifestation of
the reversal symmetry $\Theta$ and the Berry phase $\pi$ in the $\vec{Q}$
(or $\vec{q}$) space,
and robust against the band gap opening.
We propose the setup for the detection of the jump,
in which collimated waves are generated, tuned, and identified.


This work is supported by NRF (2009-0078437).



\begin{thebibliography}{99}
\bibitem{NOTE2} For bilayer graphene,
Berry phase is usually mentioned as $2 \pi$ (see, e.g., Ref.~\cite{Novoselov2}),
associated with
a $\vec{q}$-space loop encircling twice the origin.
\bibitem{NOTE1} We ignore the paths with multiple reflections at $x = d_s$ (as
$|r_1|^2 \ll 1$)
and those going back to the barrier
and re-incoming to the step. They are fully counted in Fig.~\ref{fig5:TandG}(b).
\end{thebibliography}
\end{document}